\documentstyle[aps,prl,twocolumn,epsfig]{revtex}

\def\psss#1{\begin{center}\leavevmode\hbox{\epsfxsize=2.7in\epsfysize=3.1in\epsfbox{#1}}\end{center}}

\newcommand{\be}{\begin{equation}}
\newcommand{\ee}{\end{equation}}
\def\bea {\begin{eqnarray}}
\def\eea {\end{eqnarray}}

\def\ell{i}

\begin{document}
\title{S-wave superconductivity near a surface}
\author{K. Tanaka$^{1,2}$ and F. Marsiglio$^3$}
\address{$^1$Department of Physics and Engineering Physics, 
University of Saskatchewan,\\ Saskatoon, SK, Canada S7N 5E2}
\address{$^2$Materials Science Division, 
Argonne National Laboratory,\\
9700 South Cass Avenue, Argonne, IL 60439 U.S.A.}
\address{$^3$Department of Physics, University of Alberta,
Edmonton, Alberta, Canada T6G 2J1}
\date{published in Physica C {\bf 384} (2003) 356}
\maketitle
\begin{abstract}
We study the superconducting order parameter near a surface with the
Bogoliubov-de Gennes formalism.  For definiteness we use the attractive
Hubbard model.
Near a surface, the order parameter and the density distribution exhibit 
``Friedel-like'' oscillations.  Although the local density of states 
is quite different from that in the bulk, the energy gap in the spectrum
on a surface is almost the same as the bulk value.  
In the low-density limit, however, the energy gap tends to vanish on a
surface.
\end{abstract}



\narrowtext

\section{INTRODUCTION}
\label{sec:int}

Experimental probes of various properties of superconductors can 
generally be classified into two categories -- bulk and surface probes.
Tunneling and photoelectron spectroscopy (PES) are examples in the second
category, and while PES has only recently been utilized heavily to
investigate superconductivity \cite{lynch99}, 
tunneling has long been an important tool
for determining the energy gap \cite{giaever60} and electron-phonon 
related structure \cite{rowell63} in superconductors. Through comparison
with results from bulk probes, it has been well established that the
superconducting order parameter in conventional superconductors
remains robust at surfaces.

The purpose of this paper is to investigate the extent to
which this statement is theoretically understood. In order to do so, in the
context of conventional s-wave superconductivity, we adopt a minimal
model to describe s-wave superconductivity, namely the attractive Hubbard
model. This choice allows us to explore various aspects of the problem
that otherwise might be difficult to study: electron density dependence and
weak-to-strong coupling tendencies, to name a few. The tight-binding
and lattice features of the this model also make it
amenable to numerical treatment. 
The presence of surfaces can be included in this model
simply by requiring that electrons cannot hop beyond the surfaces, e.g.,
by imposing open boundary conditions (OBC) as opposed to periodic
boundary conditions (PBC).

The paper is organized as follows. In the next section we outline
the formalism used in this study. We simply solve the
Bogoliubov-de Gennes (BdG) equations \cite{degennes66}, as described in
a previous publication \cite{tanaka00}. 
We also compare results with those obtained from the Anderson prescription
\cite{anderson59,tanaka00}, for which a brief review is provided.
In the following section we present
results for the order parameter as a function of distance from
a surface for various parameter regimes. It is clear that a large
spatial variation occurs, dependent on coupling strength and electron
density. We also examine the local density of states as a function
of distance from the surface, and find significant variation close
to the surface. 
In this work we focus our study on one dimension, 
in which the impact of a surface can be visualized clearly.
The effects of dimensionality on the ``Friedel-like'' oscillations
are discussed, and also some results for the local density of states 
are shown for a two-dimensional system.

\section{FORMULATION}
\label{sec:form}

The use of the Bogoliubov-de Gennes (BdG) equations 
\cite{degennes66,flatte99} (or a corresponding quasi-classical formulation 
\cite{rainer96}) is required to solve for the superconducting 
order parameter in cases where the order parameter varies with position.
This will generally occur in the vicinity of a defect or impurity, but
first and foremost, will always occur near a surface. Even if the surface
does not modify the material parameters, its existence causes an abrupt
change in the effective pair potential, and hence the order parameter
will vary in its vicinity.

What governs the length scale of the variation in the order parameter
as one moves away from the surface?  Three length scales enter the
problem (we leave out the actual thickness of the sample, which enters
in the Tomasch-Rowell experiment \cite{mcmillan69}, or in nanograin
superconductors). These are (i) the lattice spacing, {\it a}, (ii) the
coherence length, $\xi_0$, and (iii) a measure of the inter-electron
spacing, say, $k_F^{-1}$, where $k_F$ is the Fermi wavevector. 
Near half-filling (i) and (iii) coincide; in the low-density limit, we
will find that (iii) governs the order parameter variation. In all cases,
one can observe a `healing length', over which the order parameter
attains its bulk value. This is qualitatively described by the coherence
length, determined here with the BCS expression:
\be
\xi = \sqrt{\langle R^2\rangle}\,, 
\label{coherence_length}
\ee
where $\langle R^2\rangle$ is the mean square radius of an electron pair 
\cite{tinkham75,marsiglio90}. Strictly speaking, as defined this
length has little to do with coherence (defined for a single pair).
However, at low temperatures, to which we confine this study, this
tends to coincide with the Ginzburg-Landau coherence length, which
more properly describes the length scale of fluctuations in the order
parameter away from its mean value.

The BdG equations are \cite{flatte99,tanaka00}:
\bea
E_l u_l(\ell) = \phantom{-} \sum_{\ell^\prime}
A_{\ell \ell^\prime}u_l(\ell^\prime) + 
V_\ell u_l(\ell) + \Delta_\ell v_l(\ell)
\label{bdg_u}\\
E_l v_l(\ell) = -\sum_{\ell^\prime} 
A_{\ell \ell^\prime}v_l(\ell^\prime) -
V_\ell v_l(\ell) + \Delta^\ast_\ell u_l(\ell)
\label{bdg_v}
\eea
where
\be
A_{\ell \ell^\prime} = -t \sum_\delta \biggl(
\delta_{\ell^\prime, \ell - \delta} + \delta_{\ell^\prime, \ell + \delta}
\biggr)
-\delta_{\ell \ell^\prime} \mu\,.
\label{bdgaux}
\ee
The self-consistent potentials, $V_\ell$ and $\Delta_\ell$, are given
by
\bea
\Delta_{\ell} = |U| \sum_l u_l(\ell) v_l^\ast(\ell) (1 - 2f_l)
\phantom{aaaaaaaaa}
\label{pair_pot} \\
V_\ell = -|U| \sum_l \biggl[
|u_l(\ell)|^2 f_l + |v_l(\ell)|^2 (1 - f_l) \biggr].
\label{hartree_pot}
\eea
The index $l$ is used to label the eigenvalues (there are $2N$ of them),
the index $\ell$ to label
the sites (1 through $N$), and the composite eigenvector is
given by $\biggl({u_l \atop v_l} \biggr)$, of total length $2N$. The sums
in Eqs.~(\ref{pair_pot},\ref{hartree_pot}) are over positive 
eigenvalues only.  The $f_l$ is the Fermi function, with 
argument $\beta E_l$, where $\beta = {1/k_BT}$ with 
$T$ the temperature, and $k_B$ the Boltzmann constant. 
The single-site electron density, $n_\ell$, 
is given through Eq.~(\ref{hartree_pot}) by $V_\ell = -|U|n_\ell/2$.
As usual, the parameters $t$ and $|U|$ are the 
hopping integral and attractive coupling strength, 
$\mu$ is the chemical potential, and the subscript `$\delta$'
in Eq. (\ref{bdgaux}) refers to nearest neighbour sites in units of the
lattice spacing.

These equations are solved by iteration until full self-consistency
is achieved. If (as is often the case) a given electron density is
desired, a second iteration loop is inserted to vary the chemical
potential $\mu$, until the required average density 
$n \equiv {1 \over N} \sum_i n_i$ is attained.
The quasiparticle amplitudes are used to calculate the
local density of states (LDOS) given by
\bea
A_i(\omega)=\sum_{l}\,
\biggl[\,|u_l(i)|^2\,\delta(\omega-E_l)+|v_l(i)|^2\,\delta(\omega+E_l)\,
\biggr]\,. \label{ldos}
\eea

We will make reference to solutions to the Anderson equations in what
follows. As described in the original paper \cite{anderson59} and
explained for the attractive Hubbard model in Ref.~\cite{tanaka00}, 
these are a BCS-like set of equations for an inhomogeneous system, for
which solutions are required for the non-interacting problem.
For tight-binding systems with open boundary conditions,
these solutions can be found analytically.
Once the single-particle states are obtained, the BCS-like equations
can be solved much more easily than their BdG counterparts.
As shown previously \cite{tanaka00} and further in this work,
they provide an excellent 
qualitative description of the position dependence of the 
order parameter and the density of states, 
for a fraction of the numerical cost for BdG.

\section{RESULTS}
\label{sec:results}

\subsection{BdG vs. Anderson}
\label{subsec:anderson}

\begin{figure}
\psss{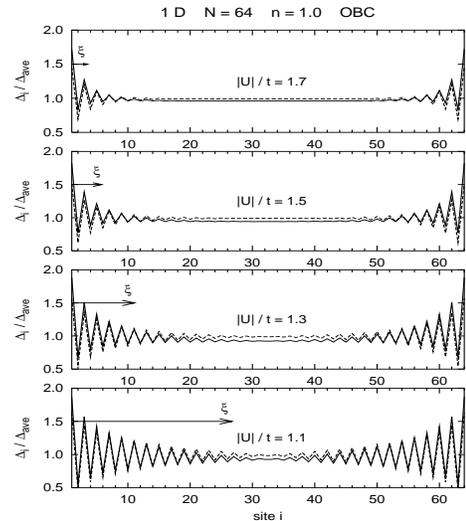}
\caption{Order parameter $\Delta_i$ normalised by the average value 
as a function of site number $i$, for a $64$-site chain with OBC 
at half filling, for various values of coupling strength $|U|$.
The BdG and Anderson results are plotted with solid and dashed curves, 
respectively, which agree very well.
The gap exhibits the ``Friedel-like'' oscillations near a surface,
which decay roughly over the coherence length $\xi$ (indicated by arrows).}
\label{fig1}
\end{figure}

\begin{figure}
\psss{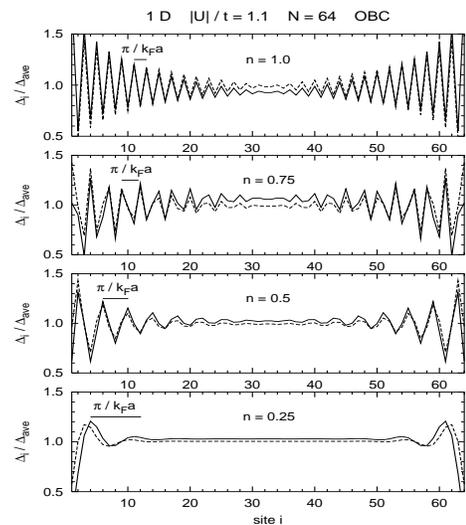}
\caption{Same as Fig.~\ref{fig1}, but for $|U|=1.1 t$ and for various 
electron densities
$n$. The period of the ``Friedel-like'' oscillations in site number $i$
is roughly $\pi / k_F a$, where $a$ is the lattice constant, and 
increases as $n$ is decreased.  
The irregular oscillations seen for $n=0.75$ arise from the incommensurability
of the electron number and the number of lattice sites.
The coherence length $\xi$ (see next figure) 
decreases as a function of decreasing $n$.}
\label{fig2}
\end{figure}

In the absence of a surface or impurity, momentum is a good quantum number,
and the BCS gap equations \cite{schrieffer64} can be solved. 
For the attractive Hubbard model, the resulting order parameter 
turns out to be constant in momentum space as well as in real space.
The presence of a surface breaks translational invariance, so that both 
the order parameter and the density distribution become dependent on position
close to the surface. In particular they exhibit ``Friedel-like'' oscillations.
In Figs.~\ref{fig1} and \ref{fig2}, 
this is demonstrated for the order parameter 
for a 64-site system with OBC, where site 1 and 64 represent the surfaces.
As mentioned above, 
the only distinction in our tight-binding formalism is that 
there is no hopping matrix element connecting the surface sites to 
outside of the chain.
Further refinements, such as altering the hopping matrix elements or the
attractive interaction for sites near the surface, are possible; however
we do not explore these possibilities here. 
Unlike the Ginzburg-Landau approaches, we do not 
impose boundary conditions on the order parameter itself:
the surface behaviour of the order parameter is
determined by the BdG equations themselves.

In Figs.~\ref{fig1} and \ref{fig2},
the BdG and Anderson results are shown with solid and dashed curves,
respectively.  Figure \ref{fig1} shows the order parameter $\Delta_i$ 
(scaled by the site-averaged value) as a function of site number $i$ for 
half filling, for various coupling strengths.
At half filling, the order parameter is peaked at the surface and 
oscillates as one
moves in towards the bulk.
These oscillations decay over the scale roughly determined by
the coherence length $\xi$, beyond which the order parameter relaxes
to its bulk value. For sufficiently large samples,
the bulk value is the same as that achieved for the homogeneous case
obtained with PBC. 
This can be seen clearly in Fig.~\ref{fig1}, with $\xi$ becoming
longer as the coupling strength is decreased.
An exception is the case with $|U|=1.1 t$, where we encounter 
finite size effects -- the system is not quite
large enough to allow relaxation of the order parameter to its bulk value.

\begin{figure}
\psss{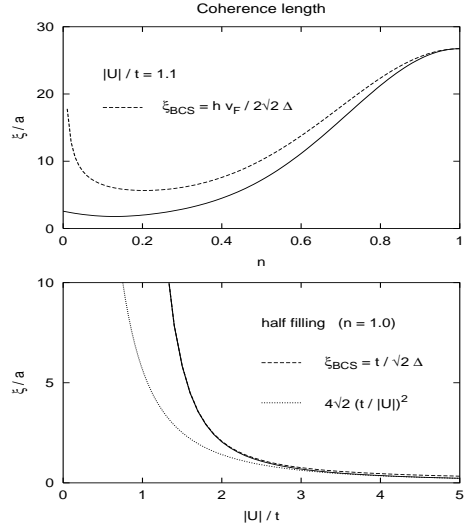}
\caption{The BCS coherence length as a function of (a) electron density, and
(b) coupling strength, with comparison to analytical approximations.
}
\label{fig3}
\end{figure}

The ``Friedel-like'' oscillations in the order parameter are a 
reflection of the single-particle wave function at the Fermi level.
For a given density $n$, the oscillation period in site number 
is given by $\pi / k_F a$, where $a$ is the
lattice spacing, and $k_F$ and $n$ are connected by the usual Pauli
consideration.
To illustrate this, we plot the order parameter with
$|U|=1.1 t$ for various values of density $n$ in Fig.~\ref{fig2}.
As $n$ decreases, $k_F$ decreases and the period becomes longer, as
is indicated by the bar in each panel.  
Note moreover that the coherence length $\xi$, for the given coupling 
strength $|U|=1.1 t$, decreases monotonically from $n=1.0$ to $0.25$:
this is shown in Fig.~\ref{fig3}(a), where we plot $\xi$ as a function of 
electron density $n$.  In Fig.~\ref{fig3}(b) $\xi$ for half filling is plotted
as a function of coupling strength $|U|$.
In these figures,
we have also included some analytical forms valid 
in the weak-coupling limit ($\xi_{BCS}$), and in (b), 
one in the strong-coupling limit as well.
Note, particularly in (b) the excellent agreement of the analytical
forms with $\xi$ in the two extremes. It is clear that in both 
Figures \ref{fig1} and \ref{fig2} the Anderson prescription (dashed
curves) captures the essential features of the ``Friedel-like'' oscillations 
in the order parameter. This is also the case for 
the electron density distribution (not shown).

\begin{figure}
\psss{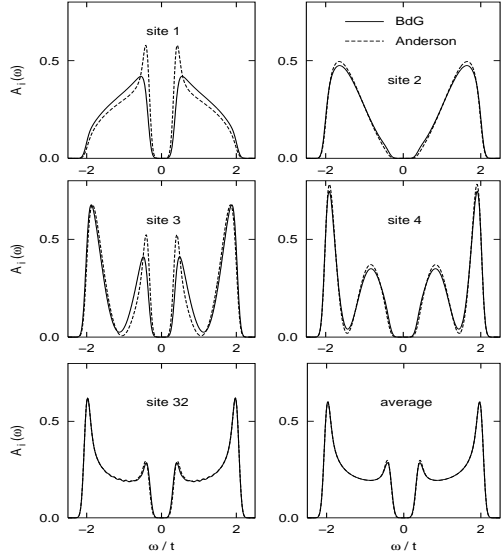}
\caption{
The LDOS for an $N=64$ system with OBC at half filling,
with $|U|/t=2.0$, for several sites: site 1 is a surface and site 32 is 
the middle of the sample. The Anderson approach (dashed curves) 
reproduces the BdG results (solid curves) remarkably well.
Near a surface, the LDOS is quite different from that in the bulk (site 32),
and at half filling, 
the energy gap is larger at every second site (e.g., site 2 and 4).
}
\label{fig4}
\end{figure}

In Fig.~\ref{fig4} we examine the local density of states calculated with
both the BdG and Anderson formalisms. 
The delta functions in Eq.~(\ref{ldos}) have each been replaced by a 
normalized Gaussian, and the smoothing width for the results shown in
all the LDOS figures is $0.1t$.
It can be seen that the Anderson prescription
reproduces the BdG results remarkably well in all details.
In this figure the LDOS at various sites is shown 
for the same system as in Fig.~\ref{fig1} but with $|U|/t = 2.0$.
Near a surface the LDOS is quite different from that in the bulk
(as represented by the LDOS in the middle of the sample, or by the average
DOS over all sites).
Although the overall LDOS as a function of energy is very different
from the bulk DOS, the energy gap in the spectrum on the surface is
almost the same as that in the bulk (compare the BdG results for site
1 and 32). 
On the other hand, it is intriguing that the energy gap is larger at every 
second site (site 2 and 4 in Fig.~\ref{fig4}): this statement applies only
to half filling. For quarter filling, for example, the energy gap is slightly
larger at every fourth site.
The Anderson results reproduce these features correctly, and their
agreement with the BdG results becomes better further away from the surface.
Note, however, the absence of coherence peaks in the more accurate BdG results
at the surface position. 

\subsection{Friedel-like oscillations}
\label{subsec:friedel}

\begin{figure}
\psss{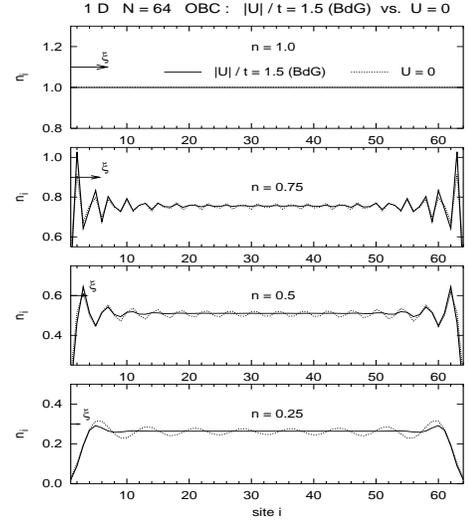}
\caption{
Density distribution $n_i$ for an $N=64$ system with OBC and $|U|/t=1.5$
for various average electron densities $n$. The BdG results (solid curves) 
are compared
with those for the noninteracting case (dotted curves).
Except for half filling, 
$n_i$ shows ``Friedel-like'' oscillations near a surface, following the
noninteracting density distribution.  
Similarly to $\Delta_i$, the length scale for the decay of these oscillations is
given by the coherence length $\xi$ (indicated by arrows).
}
\label{fig5}
\end{figure}

Next we study the ``Friedel-like'' oscillations in more detail. In 
Fig.~\ref{fig5} the density distribution $n_i$ as a function of site number $i$
is shown for a 64-site chain (OBC, $|U|/t=1.5$) for various values
of the average electron density $n$. 
In this figure the density distribution obtained
from the BdG results (solid curves) is compared with the density distribution 
for the noninteracting case (dotted curves).
Except for half filling, the density distributions in both the
superconducting and normal states exhibit oscillations, with period
determined by $k_F^{-1}$. In the superconducting state the oscillations
are smoothed as one moves away from the surface, as was the case with
the order parameter (see Fig.~\ref{fig2}). 
This distance over which the electron density 
becomes smooth is the same as for the order parameter, and is
loosely given by the coherence length $\xi$.
The exceptional case of half filling ($n=1.0$) has an electron density that
is independent of site position, due to the particle-hole symmetry 
present at half filling (see next subsection, however).

As seen in Fig.~\ref{fig5}, the density distribution (for nonzero $U$)
sometimes has higher amplitude near a surface,
particularly close to half filling.
This behaviour is not to be confused with
the decay of the ``Friedel-like'' oscillations in the order parameter.
The former simply follows the non-interacting distribution, which 
results from the particle-in-a-box problem subject to the Pauli
exclusion principle.
With OBC the single-particle states are standing waves, and
the sum of their probability distribution tends to be higher near a surface 
near half filling.

\begin{figure}
\psss{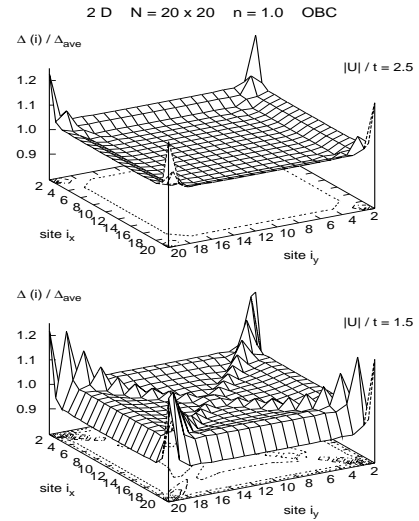}
\caption{Order parameter $\Delta_i$, as a function of position on a 2D (20X20)
lattice with OBC, for $|U|/t = 2.5$ (upper graph) and $|U|/t = 1.5$ (lower graph).
Note the variation near the surfaces and along the diagonals, particularly for
the weaker coupling case. These results are computed at half filling.
}
\label{fig2d_1}
\end{figure}

While the electron-electron interaction is local in this model, 
in two and three dimensional systems, the ``Friedel-like'' oscillations 
appear in somewhat different ways from one dimension (1D).
The difference stems mainly from two factors. One is that 
unlike in 1D, with OBC, most of the single-particle energy 
levels are degenerate in higher dimensions.  As a result, 
interference occurs among degenerate single-particle states 
at the Fermi level, and the order parameter and density distribution 
can exhibit quite complicated structure.  In two dimensions (2D)
at half filling, however, it is relatively simple due to the structure of 
the Fermi surface.  We illustrate this in Fig.~\ref{fig2d_1}, where the order
parameter is plotted for a $20\times 20$-site system for $|U|/t = 2.5$ 
and 1.5.  At half filling, the order parameter shows the 
``Friedel-like'' oscillations, where it is peaked along the
diagonal sites and largest near the corners.
As a function of decreasing $|U|$ (increasing $\xi$), the oscillations
extend towards the middle of the sample. These oscillations, as seen in 
Fig.~\ref{fig2d_1}, can be explained by the interference of 
single-particle wave functions.  In the zero-coupling limit 
($|U|\rightarrow 0$), Eq.~(\ref{pair_pot}) reduces to 
$\Delta_{\ell} \simeq |U|\,(1/2) \sum_{{\bf k}_F} u_{{\bf k}_F}(\ell) 
v_{{\bf k}_F}^\ast(\ell)$ at half filling at zero temperature, and 
$u_{{\bf k}_F}(\ell)\simeq v_{{\bf k}_F}(\ell) \simeq 
u_{{\bf k}_F}^0(\ell)$.  Here $u_{{\bf k}_F}^0(\ell)$ is the single-particle 
wave function (analytical solution \cite{tanaka00})
for the Fermi momentum ${\bf k}_F = (k_x,k_y)$, and the $\simeq$ denotes that
the oscillatory behaviour is similar. In fact the amplitude of the hole-like
part decreases with $|U|$ for electron-like states and vice-versa for the
electron-like part..
Using the relation $k_y=-k_x+\pi$ on the Fermi surface to eliminate $k_y$,
and approximating the sum by an integral over $k_x$,
we can calculate $\Delta_{\ell}$ analytically. Thus
\begin{equation}
\Delta_i \simeq \Delta\,\biggl[\,1 + {1\over 2} 
(\delta_{y_i,x_i} + \delta_{y_i,-x_i})\,\biggr],
\label{pair_weak}
\end{equation}
where $x_i,y_i$ are the $x$ and $y$ coordinates of site $i$,
and $\Delta$ is a small number.
Clearly $\Delta_{\ell}$ is larger along the diagonals ($x_i=\pm y_i$).
We have obtained the BdG solutions for $N=32^2$ at half filling 
as a function of $|U|$, and found that for very weak coupling 
(e.g., $|U|/t = 0.1$), the above equation indeed describes
the order parameter structure.

For three dimensional systems, the largest size we have examined is
$N=12^3$.  For coupling strengths small enough to
see the ``Friedel-like'' oscillations, however, this size is still small
-- there are only 12 sites on one side and the surfaces occupy a large
portion of the system -- so that the order parameter is dominated by 
finite size effects.  We have calculated the three-dimensional version of
Eq.~(\ref{pair_weak}) numerically, and found that it qualitatively
explains the oscillatory structure of the order parameter in the weak-coupling
limit.

\begin{figure}
\psss{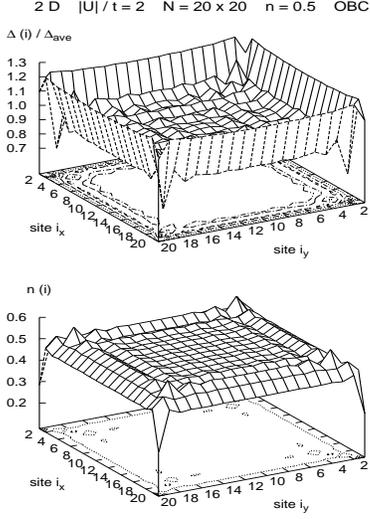}
\caption{Order parameter $\Delta_i$ (upper graph) and $n(i)$ (lower graph), 
as a function of position on a 2D ($20\times 20$) lattice with OBC, 
for $|U|/t = 2$. These results are computed at quarter filling
($n = 0.5$).  Note the variation near the surfaces.
}
\label{fig2d_2}
\end{figure}

As $n$ becomes smaller than 1, the interference pattern changes. As seen above,
in 2D, for $n\simeq 1$, the Fermi surface is approximately $k_y=-k_x+\pi$
and the oscillations in $\Delta_{\ell}$ as well as $n_i$ are along 
several sites around the diagonal. Away from half filling, the oscillations
occur everywhere along the surfaces.
Figure \ref{fig2d_2} shows the order parameter and density distribution
for $N=20^2$ and $|U|/t = 2$ at quarter filling.  
In the low-density limit, both $n_i$ and $\Delta_{\ell}$ are suppressed
on and near the surfaces.

\begin{figure}
\psss{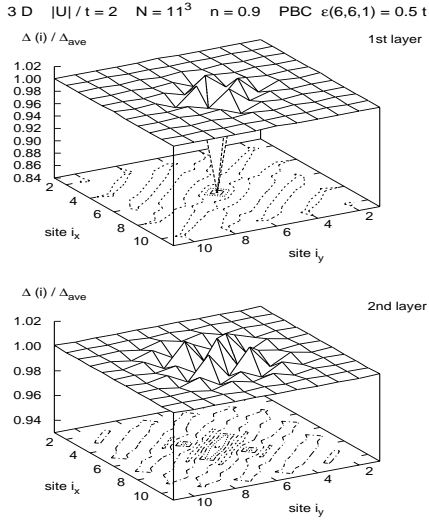}
\caption{Order parameter $\Delta_i$, as a function of position on a 3D 
($11\times 11\times 11$) lattice with PBC, for $|U|/t = 2$, 
with the average electron density, $n = 0.9$.
A single-site impurity is located in the middle of the surface layer (upper
graph) with strength $\epsilon = 0.5t$. The variation of the order parameter is
also shown in the second layer (lower graph). Note the condensed scale in the 
lower graph, indicating that the disruption of the order parameter is 
concentrated near the impurity.
}
\label{fig3d_1}
\end{figure}

Another factor that distinguishes one dimension from higher dimensions
is the dependency of nesting on dimensionality.  
In 1D, the Fermi surface is perfectly nested
for all values of $n$.  In 2D and 3D this is the case only at half filling,
and this means that away from half filling, ``Friedel-like'' oscillations
can be suppressed compared with 1D.  
We have not identified any noticeable difference in this sense between
1D and 2D.  For 3D, we have examined this aspect by placing a
nonmagnetic impurity which shifts the chemical potential at that site 
and using PBC.  In this way a disturbance that breaks translational
invariance is one site, as opposed to all the surfaces.  
In Fig.~\ref{fig3d_1} we show the order parameter for $N=11^3$, $n=0.9$,
and $|U|/t = 2$, in the first layer (top) and the second layer (bottom).
The impurity potential is $0.5t$ in the middle of the first layer.
The basic features caused by an impurity are the same as in 1D
\cite{tanaka00}: in the first layer 
$\Delta_{\ell}$  is suppressed at the impurity site
and shows the ``Friedel-like'' oscillations around it.  The size-dependent
electron density, $n_i$
(not shown), behaves similarly, except that for an attractive potential,
it is peaked at the impurity site.  The $\Delta_{\ell}$ is peaked
right below the impurity site in the second layer and suppressed in the
third layer, and so on.  However, the extent to which the oscillations 
propagate is smaller compared with 1D and 2D.
This is the case within the layer where the impurity is, as seen in
Fig.~\ref{fig3d_1}.  Moreover,
the oscillation amplitudes decrease quickly as one moves away from that
layer: note that in Fig.~\ref{fig3d_1} the $z$-axis range 
for the second layer is half of that for the first layer.

\begin{figure}
\psss{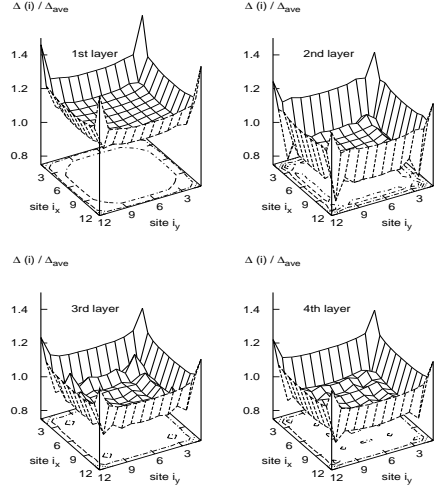}
\caption{Order parameter $\Delta_i$, as a function of position on a 3D 
($12\times 12\times 12$) lattice with OBC, for $|U|/t = 2$, at half filling, 
$n = 1$. Results are shown for
four layers, starting with the surface layer. The order parameter displays
considerable variation near all the surfaces.
}
\label{fig3d_2}
\end{figure}

In the strong-coupling limit, at half filling in any dimension,
the order parameter is larger on the surfaces (largest at the corners) and 
more or less flat inside.  In Fig.~\ref{fig3d_2} we show $\Delta_{\ell}$
for $N=12^3$ and $|U|/t = 2$ at half filling, for the first four layers.
While $\Delta_{\ell}$ is larger on the surfaces (the first layer and
peripherals of each layer), some oscillations near the surfaces can be seen.

\subsection{CDW}
\label{subsec:cdw}
 
For the half-filled attractive Hubbard model with nearest neighbour hopping
on a translationally invariant system,
the superconducting (SC) and charge-density-wave (CDW) states are degenerate. 
The presence of an impurity tends to make a CDW state more favourable,
and one might expect that a surface has the same effect. This is not
necessarily the case, however.  With an even number of sites $N$, with
OBC, the SC and CDW states are degenerate at half filling. Moreover, the
state with constant density distribution discussed above is not the
only SC solution. When the density distribution of a CDW state is used as
input, the BdG equations converge to a SC state with a similar density
distribution, and this state has the same energy as the one with constant
density.  The order parameter structure is the same for both states, but
the overall magnitude is smaller for the former.  For any average density,
the BdG equations always converge to a SC solution.
 
The situation is different, however, for an odd-$N$ system with surfaces.
The BdG equations converge to a SC solution at any density that
corresponds to an odd number of electrons, e.g., at half filling. For an
even number of electrons, the BdG equations either do not converge at all,
or converge to a CDW solution. This presumably is caused by the 
incommensurability
of the electron number (all paired) and the number of sites. The same CDW
solution for an even number of electrons is obtained by solving the
Hartree equations (no superconducting order parameter).
With $N+1$ electrons the system is equivalent to an $N+1$ 
system with PBC at half filling, with a strong repulsive impurity at one site
where the electron density becomes zero.
While the Hartree equations usually do not converge for an odd electron
number, we were able to find a CDW solution at half filling with odd $N$.
It is interesting to note that unlike even $N$, the CDW state has a lower
energy than the SC state, and we did not find a SC state with an
oscillating density distribution.
 
\begin{figure}
\psss{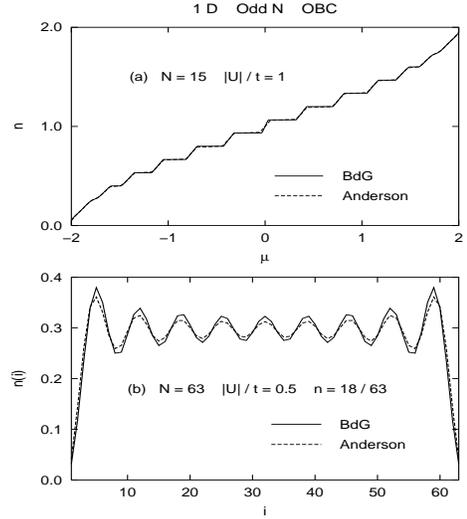}
\caption{(a) Average electron density vs. chemical potential and 
(b) electron density as a function of position, for an odd number of sites
$N$ with OBC.
Solid curves (dashed curves) are BdG (Anderson) results. 
We used $|U| = 1t$ and $N = 15$ in (a) and $|U| = 0.5t$, $N = 63$, 
and $n = 18/63$ in (b).  The latter graph is for a CDW state (with zero
superconducting order parameter).
}
\label{cdw}
\end{figure}

In Fig.~\ref{cdw}(a) we show the average density $n$ vs. chemical
potential (``dressed'' with the Hartree energy), obtained by BdG for 1D with
$N=15$ with $|U|/t = 1$. The plateaus correspond to CDW states, where the
superconducting order parameter is zero.  For stronger coupling and larger
system size, the plateaus tend to disappear, but the BdG equations pick up
the CDW solutions at certain densities.
In Fig.~\ref{cdw}(b) the density distribution is shown for a CDW state
in 1D with $N=63$, $|U|/t = 0.5$, and $n=18/63$. 
In both (a) and (b), the BdG and Anderson
results are plotted with solid and dashed curves, respectively.
Although the Anderson results are similar to the BdG results, they are not
CDW solutions: the density distribution is the same as the one in the
noninteracting case.

\subsection{Local density of states}
\label{subsec:ldos}

\begin{figure}
\psss{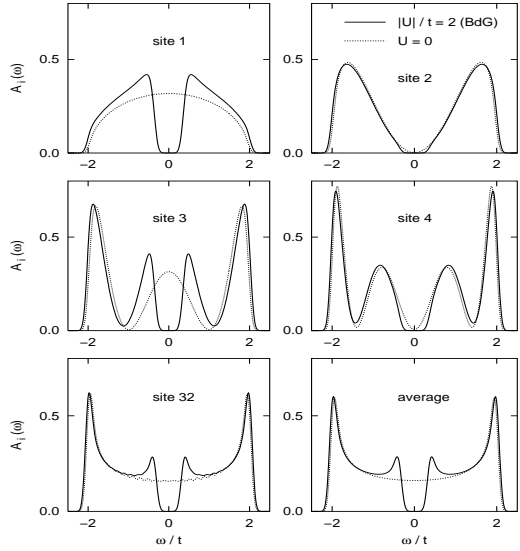}
\caption{
Same as in Fig.~\ref{fig4}, but the BdG results (solid curves) are compared 
with those for the noninteracting case (dotted curves).
}
\label{fig6}
\end{figure}

To distinguish the features that are due to the opening of a superconducting
gap from those that are due to the single-particle states, we compare, in
Fig.~\ref{fig6}, the LDOS (the same BdG results as shown in Fig.~\ref{fig4}) 
with that in the noninteracting case, for various sites near the surface.
Interestingly, the noninteracting LDOS has a gap-like feature
at every second site (e.g., site 2 and 4) near a surface (again this is
the case only for half filling). This is consistent
with a larger gap at these sites in the superconducting state.

As seen in Figs.~\ref{fig4} and \ref{fig6}, the energy gap as it appears in 
the LDOS on the
surface is not very different from the one in the bulk, despite the fact that
both the order parameter and the density distribution behave quite 
differently on and near the surface. However, the BCS coherence factors,
which give rise to peaks in the density of states just beyond the gap
(as seen in the LDOS for site 32 in Fig.~\ref{fig6}), are barely existent
near the surface; for half filling, particularly at the even-numbered sites. 
For an orbital structure
much more complicated than the single-band one-dimensional model adopted
here, these effects would have to be considered when interpreting STM or
ARPES data.

\begin{figure}
\psss{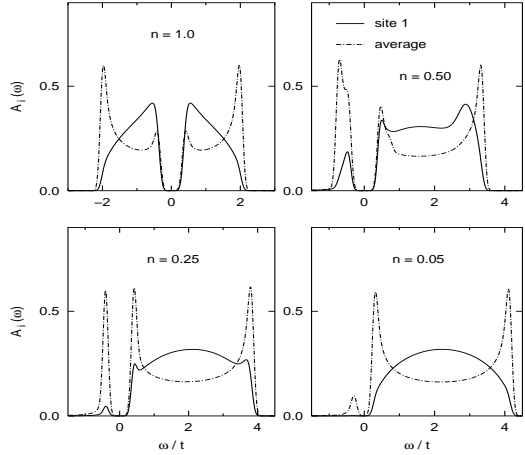}
\caption{
The LDOS for an $N=64$ chain with OBC and $|U|/t=2.0$,
for several average electron densities $n$, obtained by self-consistent 
solution to 
the BdG equations.
Those at the surface (solid curves) are compared with the average for
all sites (dash-dotted curves), which has the same behaviour as the bulk DOS.
The former differ markedly from the latter, particularly in the 
low-density limit.
}
\label{fig7}
\end{figure}

The site variation of the LDOS also depends on electron density.
In Fig.~\ref{fig7} we plot the LDOS at the surface of a 64-site chain for
various average electron densities. 
We have solved the BdG equations self-consistently
for $|U|/t=2.0$ and for the given values of electron density. The resulting
LDOS is compared to that calculated in the bulk (we simply averaged 
over all sites, which gives a result identical to that obtained in 
the interior, or that obtained with PBC).
Particularly in the low-density limit large differences
occur: the bulk DOS always has a well-defined energy gap, with remnants of
the BCS coherence peaks, whereas the LDOS on the surface does not. 
This occurs mainly because in the low-density limit,
the electron density at the surface sites is
significantly lower than in the bulk.
This is most apparent in Fig.~\ref{fig5}, where the electron density 
({\it even in the non-interacting limit}) is considerably lower near the 
surface than in the
bulk. This is true even when we include repulsive Coulomb interactions
between electrons, as we have verified by direct calculation in the 
low-density limit. In other words,  
Coulomb interactions do not affect the inhomogeneous electron
distribution as much as one might think because we are in the dilute limit.

\begin{figure}
\psss{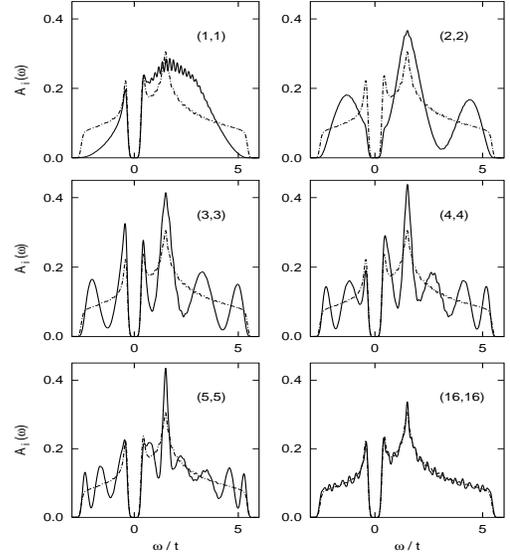}
\caption{Site dependent LDOS for $N = 32^2$ and $|U|/t = 2.5$ at quarter 
filling ($n = 0.5$), for several sites near the corner (along the diagonal). 
The dot-dashed curves are the average (bulk) result. 
}
\label{ldos2d}
\end{figure}

In 2D and 3D, the LDOS behaves in a similar way as in 1D, as one moves away
from a corner along the diagonal.
In Fig.~\ref{ldos2d} the LDOS for $N=32^2$ and $|U|/t=2.5$ at quarter filling 
is shown for several sites near the surfaces.

\section{SUMMARY}
\label{sec:concl}

We have calculated the spatially inhomogeneous order parameter
and local density of states (LDOS) for a very simple model 
exhibiting s-wave superconductivity, as a function
of coupling strength and electron density. 
We have used the Bogoliubov-de Gennes
formalism and solved the resulting equations self-consistently, including
both the anomalous and normal effective potentials. We find that both
the order parameter and the density distribution 
vary significantly as a function
of position near the surface. The details of this variation are dependent
on both the electron density and the coupling strength. The LDOS shows a
similar variation. If one can somehow preferentially tunnel electrons
into the surface
or subsurface layers, this variation may be observable \cite{trugman}. 
Here again, details
will be dependent on the specifics of the model, i.e., the character of the
orbitals involved in the superconductivity as well as the symmetry of
the superconducting order parameter. We intend to further investigate these
issues in future work.

\acknowledgments

We thank Matthias Eschrig and Stuart Trugman for stimulating discussions.
This research was supported by the
Natural Sciences and Engineering Research Council of Canada and
the Canadian Institute for Advanced Research.

\pagebreak

\end{document}